\documentclass[10pt,final,journal,twocolumn]{IEEEtran}
\usepackage[utf8]{inputenc}
\usepackage{cite}
\usepackage[cmex10]{amsmath}
\usepackage{amsmath}
\usepackage{amsfonts}
\usepackage{amsthm}
\usepackage{algorithm}
\usepackage{multirow}
\usepackage{multicol}
\usepackage{stfloats}
\usepackage{multicol,multienum}
\usepackage{multirow}
\usepackage[caption=false,font=footnotesize]{subfig}
\usepackage{wrapfig}
\usepackage{color}
\usepackage{graphicx}
\usepackage{xspace}

\def\etc{\textit{etc.}\xspace}

\begin{document}

\title{Explainable Artificial Intelligence (XAI) for 6G: Improving Trust between Human and Machine
\thanks{Weisi Guo is with the Alan Turing Institute, London, United Kingdom, and Cranfield University, Bedford, United Kingdom. $^*$Corresponding Author: wguo@turing.ac.uk.}
}

\author{
\IEEEauthorblockN{Weisi Guo \textit{IEEE Senior Member, RSS Fellow}}}

\maketitle

\begin{abstract}
As the 5th Generation (5G) mobile networks are bringing about global societal benefits, the design phase for the 6th Generation (6G) has started. 6G will need to enable greater levels of autonomy, improve human machine interfacing, and achieve deep connectivity in more diverse environments. The need for increased explainability to enable trust is critical for 6G as it manages a wide range of mission critical services (e.g. autonomous driving) to safety critical tasks (e.g. remote surgery). As we migrate from traditional model-based optimisation to deep learning, the trust we have in our optimisation modules decrease. This loss of trust means we cannot understand the impact of: 1) poor/bias/malicious data, and 2) neural network design on decisions; nor can we explain to the engineer or the public the network's actions. In this review, we outline the core concepts of Explainable Artificial Intelligence (XAI) for 6G, including: public and legal motivations, definitions of explainability, performance vs. explainability trade-offs, methods to improve explainability, and frameworks to incorporate XAI into future wireless systems. Our review is grounded in cases studies for both PHY and MAC layer optimisation, and provide the community with an important research area to embark upon.
\end{abstract}


\IEEEpeerreviewmaketitle

\section{Introduction}

An essential fabric of modern civilization is the digital economy, which is underpinned by wireless communication networks. We are on the cusp of entering a new era of mass digital connectivity, where increasingly more people, machines, and things are being connected to automate and digitise traditional services. Wireless networking has transitioned from its traditional role as an information channel (1G to 3G) to a critical leaver in the new industrial revolution (5G and beyond to 6G \cite{6G}). This has caused not only up to 1000$\times$ growth in the communication data rate demand, but also an increase in diverse service requirements, such as massive URLLC for tactile control of autonomous entities across transport to precision manufacturing in 6G. Orchestrating co-existence via spectrum aggregation between different radio access technologies (RATs) is essential to meeting this demand. As such, real-time radio resource management (RRM) is critically important, but has become too complex for conventional optimisation. This brings the need to evolve towards an Artificial Intelligence (AI) driven ecosystem \cite{AI5G} to support more fine-grained user-centric service provision (see 3GPP Release 16 TR37.816). Research on the application of machine learning in 5G PHY and MAC layers can be found in IEEE ComSoc Best Readings in Machine Learning in Communications \footnote{https://www.comsoc.org/publications/best-readings/machine-learning-communications}. 

\subsection{AI and Trust}
An open challenge with Deep Learning (DL) is the lack of transparency and trust compared to traditional model-based optimisation. Neural networks (NN), especially when coupled with reinforcement learning (e.g. deep reinforcement learning - DRL \cite{DRL_application_survey}) cannot explain the essential features that influence actions, nor the impact of bias on the uncertainty of rewards. This is made harder in high-dimensional mobility scenarios, such as joint airborne UAV and ground vehicle environments \cite{UAV-AI3}. Even in a relatively trusted area of Bayesian inference, recent research have shown that they are extremely brittle to poor data. As such, there is the need to develop statistical AI algorithms that can quantify uncertainty, especially mapping big data inputs, algorithm design, to the projected wireless key performance indicators (KPI). A trustworthy AI should be able to explain its decisions in some way that human experts can understand (e.g. the underlying data evidence and causal reasoning). Understanding both our opportunity and vulnerability to AI and big data is essential to the success of future customised wireless services. 

\subsection{Novelty \& Organisation}

In this review, we outline the core concepts of Explainable Artificial Intelligence (XAI) for 6G, including the key novelties in their corresponding sections: 
\begin{enumerate}
    \item Section II-A: Public and legal motivations for improving the transparency and trust in AI algorithms;
    \item Section II-B: Definitions of explainability from specific quantitative indicators, to general qualitative outputs;
    \item Section III: Review of current deep learning techniques in PHY and MAC layer and their level of performance vs. explainability trade-off;
    \item Section IV: Technical methods to improve explainability in deep and deep reinforcement learning methods;
    \item Section V: Propose an encompassing framework to incorporate XAI into future 6G wireless systems;
\end{enumerate}
Our review is grounded in cases studies for both PHY and MAC layer optimisation, including examples of explainability in existing algorithms. Together, the author hope this article provide the community with an important research area to embark upon. 

\section{Motivation and Definitions of XAI}
\label{section2}

\subsection{Public Trust \& Legal Frameworks}

At the heart of our need to add explainability/interpretability/openness to deep learning is the need to build trust in a quantifiable way. Traditional model based techniques have reasonably high clarity in how an \textit{assumed model} and the \textit{input data} leads to \textit{output decisions}, i.e., Bayesian inference gives a statistically sound framework for mapping the confidence in our data to the model outcomes. However, deep learning (DL), at least in its naive form, has none of the above. It's infamous "black box" approach yields strong performance due to the automated discovery of high-dimensional nonlinear mappings, but human operators cannot understand the following:
\begin{itemize}
    \item What data features are contributing to decisions and where do we need more or better data
    \item How to improve the algorithm design, when more and better data is not helping
    \item Uncover hidden bias in both the input data and the algorithm
    \item Reverse teach human experts to uncover new insights
\end{itemize}
Not every XAI method will address all of the above, and this paper sets out the methodology for addressing some of the above challenges.

The \textit{legal framework} for AI is still in its infancy, and there are several explicit cases for XAI in different geographic regions: 
\begin{itemize}
    \item EU: 2018 General Data Protection Regulation (GDPR) in EU requires machine learning algorithms to be able to explain their decisions (see Recital 71). 
    \item USA: 2017 Equal Credit Opportunity Act update (Reg B, art 1002.9) requires agencies to provide an official set of reasons on the main factors that affect the credit score.
    \item National: 2016 French Digital Republic Act requires the degree and mode of algorithms that contribute to decisions, the data used and its provenance, the weight of different data features, and the resulting actions. 
\end{itemize}
There is ongoing debate on whether there is a negative bias towards machine decisions (when humans do not always need to explain their actions). The key is that rightly or wrongly, humans can attempt to explain if prompted to, and we need machines to have that equal capability in order to ensure trust and a concrete pathway towards improving safety and reliability.

\subsection{Definitions and Modes of Explainability}

The essential functional role of that machine learning plays in wireless systems has not changed compared to classic model based techniques. In one way or another, the mathematical representation is:
\begin{equation}
    \hat{y} = f(x),
    \label{ML_Model}
\end{equation} where inputs $x$ map to an estimated output of true $y$ ($\hat{y}$) via a model $f(\cdot)$. In classic statistics, we apply Bayesian inference to estimate the parameter values of a known function, e.g. $\theta$ in the example linear mapping of $\hat{y} = \theta x + n$. In DL, we automate the discovery of nonlinear mapping between input and output via the training process. 

An intuitive and good starting point for \textit{explainability} is for it to meet two conditions:
\begin{enumerate}
    \item Prediction is correct, e.g. $\hat{y} = y$, and
    \item Prediction is based on the correct data features and logic, e.g. aspects of $x$ combined with the form of $f(\cdot)$ are agreeable to human reasoning/experience.
\end{enumerate}
The latter is much harder to define, let alone articulate in a DL framework. This is particularly challenging when we are dealing with DRL, large input data sets, and multiple hidden layers -- we will discuss these aspects later in the paper. For now, we discuss the different modes of explainability that we may wish for or can only have. 

\subsubsection{Visualisation with Case Studies}
The simplest form are visual outputs from the DL algorithm highlighting features in raw data that causally lead to the output choice. This may or may not map to the human perceptions of key features which also contribute to our cognitive reasoning. When combined with well known case studies, whereby the input and output mapping is established, we can both satisfy that predictions are correct and it is likely the human operator can easily accept or reject the key visual features.  

\subsubsection{Hypothesis Testing}
A more rigorous form of the aforementioned is hypothesis testing, whereby a well formulated argument is tested based on the input data and output decision. Here, we can test if: i) certain key features are important in the mapping, ii) the mapping function behaves as we expect (monotonic, nonlinear, ...etc.), and iii) we can accept or reject the hypothesis.

\subsubsection{Didactic Statements}
Perhaps the ultimate form of explainability would use natural language to communicate to the human operator, explaining what data features and algorithmic functions led it to reach a decision/output. This requires very strong explainability, as well as a machine-human interface to explain the learning and decision process. 

Now that we have established our motivation for understanding deep learning from a human reasoning perspective, and the ways in which this might be measured and manifested, we jump deeper into the wireless context to see to what degree this can be accomplished. 

\begin{figure*}[t]
     \centering
     \includegraphics[width=0.9\linewidth]{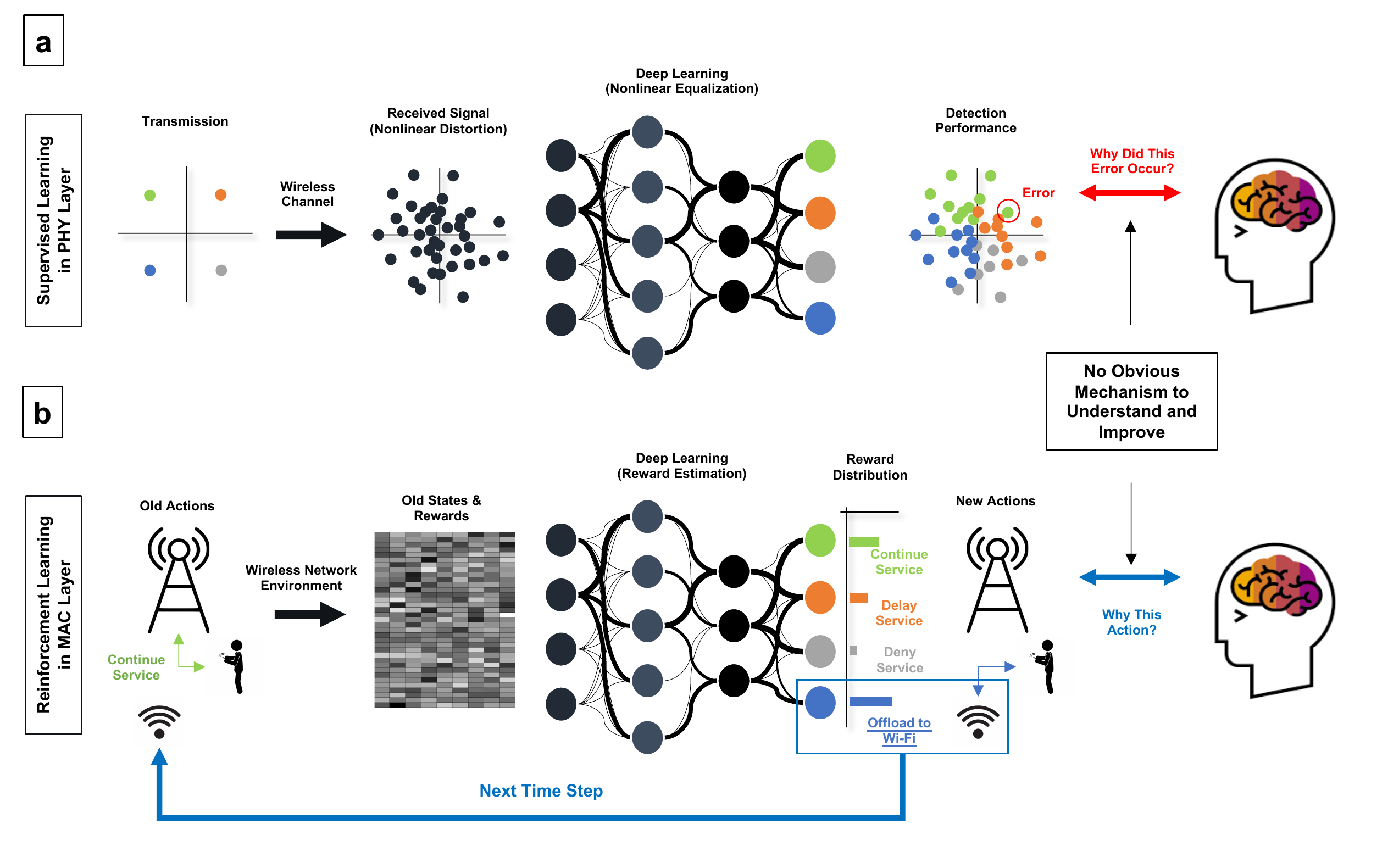}
     \caption{Examples of deep learning applications in PHY and MAC layers: a) supervised equalisation of nonlinear symbol distortion; b) reinforcement learning of action choice in offloading users.}
     \label{fig1}
\end{figure*}

\begin{table*}[!t]
\renewcommand{\arraystretch}{1.2}
\centering
\caption{AI Examples in Wireless Communication}\vspace{-1mm}
\begin{tabular}{|c||c|c|c|c|c|}
\hline
\bfseries Problem Domain &  \bfseries Representative Paper &  \bfseries Classic Approach &  \bfseries AI or DL Approach & \bfseries Improvement at BER & \bfseries Explainability \\
\hline\hline
Signal Detection & Ye18 (WCL)  & DFT with LS or MMSE & DNN with 3 hidden & $>$15dB at $10^{-1}$ & Low  \\ 
Channel with Memory & Farsad18 (TSP)  & Viterbi Detector (VD) & SBRNN with 1 hidden & 20 VD mem. at $10^{-1}$ & Low  \\ \hline
Decoding of LDPC & Nachmani18 (JSTSP) & Belief Propagation (BP) & RNN with 5 hidden & 1dB at $10^{-3}$ & V. Low \\ \hline
Channel Estimation & Neumann18 (TSP) & Orth. Matching Pursuit & CNN with 1 hidden & 2dB at $10^{-1}$ & Low \\
NOMA SCMA Detection & Kim18 (CL) & Message Passing & DNN with 4 hidden & 2dB at $10^{-3}$ & V. Low \\
Channel Est. mm-M-MIMO & He18 (WCL) & Support Detection & CNN and 3 layers & 17dB at 5dB SNR & V. Low \\ \hline
Cognitive Radio & Tsakmalis18 (JSTSP) & Expectation Prop. & Bayesian MCMC & 25 flops at $10^{-1}$ error & Medium \\
Power Allocation & Nasir19 (JSAC) & Frac. Prog. \& WMMSE & DQN`with 3 hidden & 1bps/Hz & None \\
Cross RAT Channel Access & Yu19 (JSAC) & RL & DQN with 6 hidden & 5\% rate & None \\
Interf. Align with Cache & He17 (TVT) & RL & DQN with 4 hidden & 20\% rate & None \\ \hline
Antenna Sel. & Joung16 (CL) & MaxMinNorm & SVM & 5\% at $10^{-1}$ & Low \\ \hline
WSN Diagnostics & Liu10 (TON) & Clustering & Bayesian Belief Net. & 5\% & Medium \\ \hline
User Behaviour Recog. & Wang10 (TMC) & SVM & Random Forest & 2-6\% & Low \\ \hline
QoE of Multimedia & Hameed16 (TM) & Fixed & Decision Tree & 50\% overhead & High \\ \hline
\hline
\end{tabular}
\vspace{-2mm}
\label{Tab1}
\end{table*}

\section{Deep Learning in Wireless: Explainability vs. Performance}
\label{section3}

\subsection{Review of Deep Learning \& Wireless Applications}

\subsubsection{PHY Layer}
Supervised DL has a wide range of applications in the PHY layer. In signal detection, it can equalise non-linear distortions by feeding the received signals corresponding to transmit data and pilots \cite{Li18}, outperforming classic MMSE approaches - see example in Fig.~\ref{fig1}a. When channels have memory, a bidirectional recurrent neural network (RNN) is more suitable and does not require channel state information (CSI), out performing Viterbi detection \cite{Farsad18}. Similar approaches for block code decoding, channel estimation for mm-Wave Massive MIMO, and end-to-end channel estimation have also been performed -- a summary of their performances is given in Table~\ref{Tab1}, along with their reported performances and potential level of explainability. 

\begin{figure}[t]
     \centering
     \includegraphics[width=1.0\linewidth]{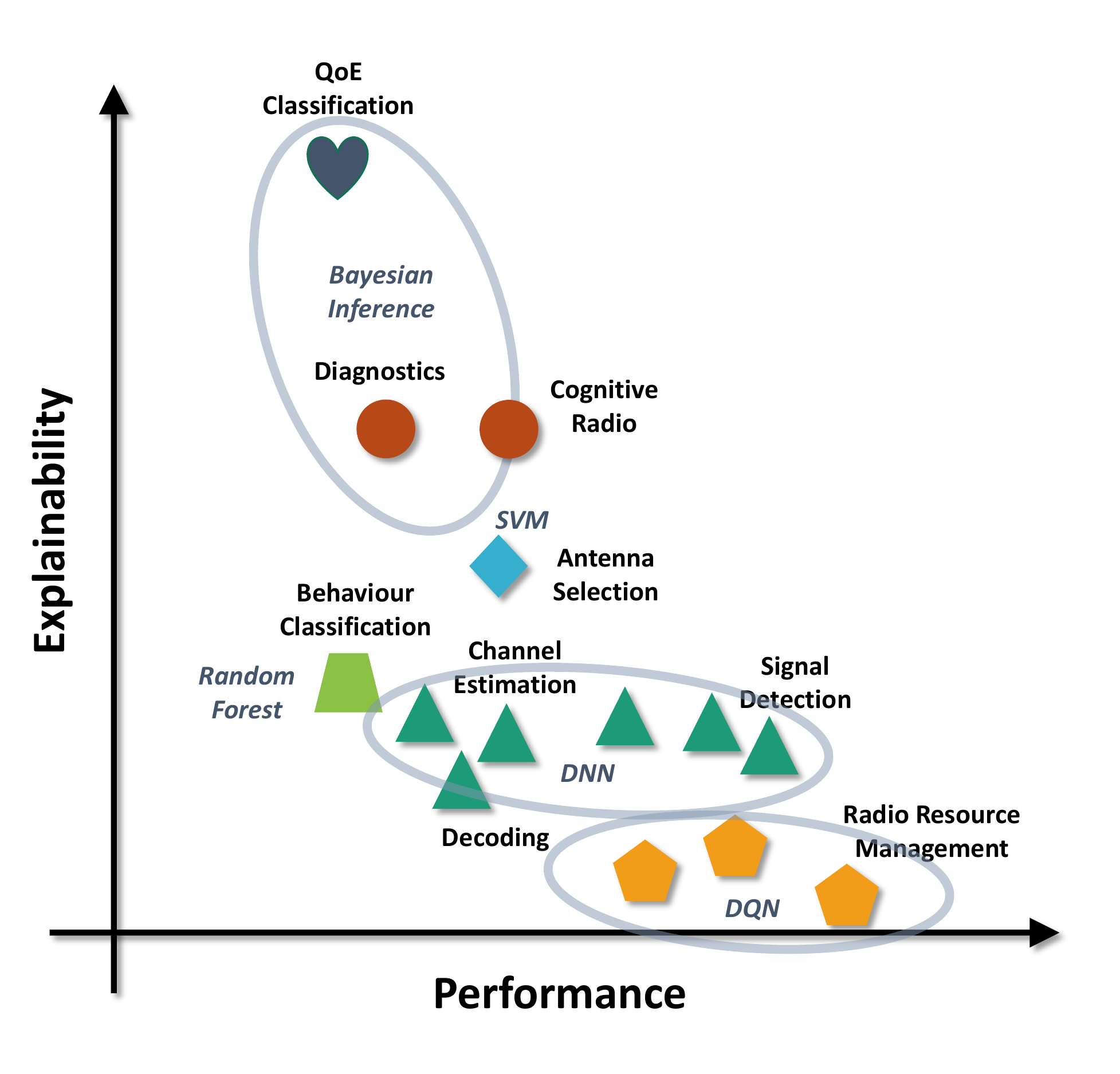}
     \caption{Trade-off between AI performance and explainability with a variety of PHY and MAC layer examples.}
     \label{fig2}
\end{figure}

\subsubsection{MAC Layer}
In MAC layer RRM, classic reinforcement learning (RL) based solutions do not rely on accurate system models and is able to run in a model-free manner. Whilst this overcame the challenges faced by traditional model dependent optimisation (e.g. dynamic programming and convex optimisation), the Q-table used in RL cannot scale to more complex problem sets such as coordinated BS offloading to heterogeneous devices, and will lead to non-convergence and high cost. Deep RL (DRL \cite{DRL_application_survey}) relies on the powerful function approximation and representation learning properties of DNN to empower traditional RL with robust and high efficient learning. In Fig.~\ref{fig1}b, we demonstrate an example of offloading user traffic based on observed state (e.g. interference, load, signal strength,...etc.), and reward (e.g. spectrum efficiency, energy efficiency) inputs. This in turn is translated into a reward distribution over possible actions (e.g. continue service, offload to WiFi,...etc.) and an action is selected. In the next time iteration, the consequence of those actions are observed. There has been a number of papers \cite{DRL_survey} that have examined the use of DRL in cellular communications, including in relatively complex mobility settings \cite{UAV-AI3}. We will not exhaustively list them here, but we will review their performance and explainability trade-off below. A summary of their performances is given in Table~\ref{Tab1}, along with their reported performances and potential level of explainability. Currently, most existing DRL solutions applied in RRM use off-the-shelf algorithms with little consideration on the RRM feature set and DRL design. This means that the resulting benefit and penalties incurred (e.g. latency and energy consumption) cannot be understood by the radio engineers monitoring and configuring the network. In order to achieve a trusted autonomy, the DRL agents have to be able to explain its actions for transparent human-machine interrogation.

\subsection{Trade-off Mapping and Interpretation Bias}

In Fig.~\ref{fig2} we show a generalised mapping of AI algorithms reviewed in Table~\ref{Tab1}. Here, we can see that Bayesian techniques (of which decision trees can also fit into) have a high degree of explainability, mapping data evidence to model form to parameter estimation and output confidence. 

Even when Bayesian inference is problematic, we tend to understand why \cite{Brittle}, e.g. when:
\begin{enumerate}
    \item the number of outcomes is large, e.g. higher order modulation (64-QAM) or continuous actions (power control levels even when discretized)
    \item a large number of marginals of the data-generating distribution are unknown (e.g. unknown mobility speed distribution amongst a range of autonomous vehicles)
\end{enumerate}
We also know how this affects AI decisions: (i) two sets of data from the same situation may appear completely different and lead to different decisions, or (ii) small changes in the model or data (its prior) can cause a different posterior conclusion. We detail more on data and algorithm bias below.

As we move away from the Bayesian framework, non-linear classification techniques such as Support Vector Machine (SVM) and random forest (RF) quickly lose explainability and there is no clear reason why data leads to one type of classification nor do we understand how over-fitted it is. For example, RF finds the optimal decision tree, but is often vulnerable to random permutation in out-of-bag (OOB) samples, otherwise known as Mean Decreased in Accuracy (MDA). The problem of sample bias and overfitting is further exasperated when we use DL to resolve a wide range of signal detection and channel estimation problems. As we wrap a RL framework around DL, we further complicate the explainability model, reaching almost zero explainability in the DRL naive form.

\begin{figure}[t]
     \centering
     \includegraphics[width=1.0\linewidth]{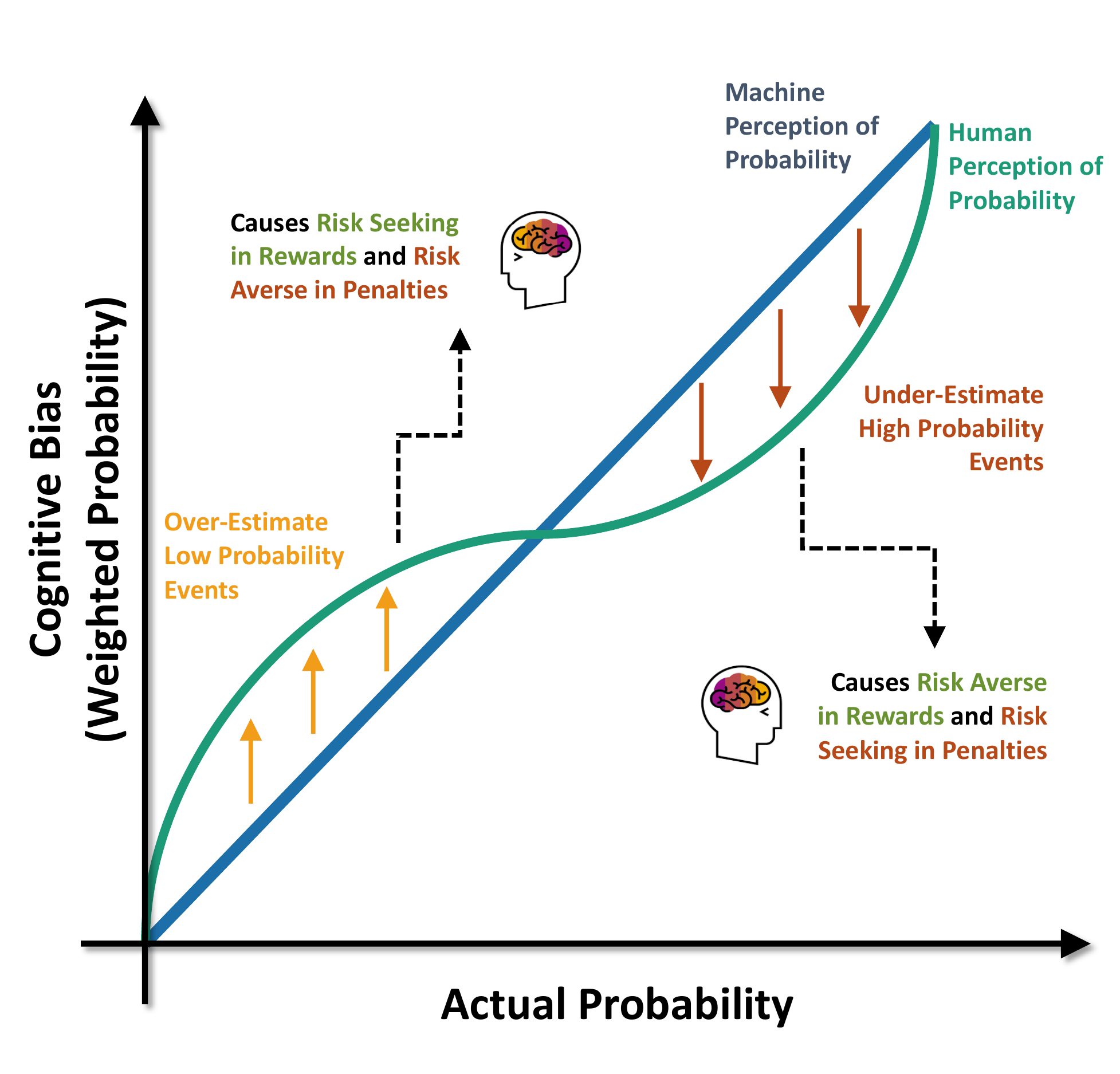}
     \caption{Cognitive bias in interpreting probability: human bias under-estimate high probability and over-estimate low probability events.}
     \label{fig4}
\end{figure}

Whilst its classification performance in complex problems is superior to the aforementioned Bayesian and classic non-linear techniques, it doesn't perform well for simple problems nor when there is clear bias. Bias in DL is not as well documented. First, it maybe intuitive to think that the weights connecting units may reveal insight (partial explainability) to its high performance -- indeed we show this is the case for many problems below. However, in some experiments it has been shown that random linear combinations of high level units also perform well. This leads to the second well known observation, which is that DNNs learn mapping $f(\cdot)$ in a discontinuous way. As such, adding purposefully designed input data noise (with no explainable features) into a well established classifier can lead to severe mis-classification \cite{ICLR14}. This remains an open challenge which we discuss more at the end of the paper.

Even if machine intelligence can explain the probability of rewards and penalties in reinforcement learning, there is a risk that humans will not perceive it in the same way as the machine. It is well known in \textit{Prospect Theory} (2002 Nobel Memorial Prize in Economics) that we have a cognitive bias in interpreting probability for rewards and penalties. Whilst machine utility functions are based on logic, human cognitive bias tend to under-estimate high probability events and over-estimate low probability events. Reliability aside, this leads us to prefer to \textit{avoid risks} in high probability reward situations (e.g. 100\% chance of 100 Mbps $>$ 90\% of 120 Mbps), and \textit{risk seeking} in high probability penalty situations. Conversely, it also leads us to prefer to seek risks in low probability reward situations, and risk adverse in low probability penalty situations. As such, not only should machines reduce bias, but must account for this human-machine difference in interpretability. This area of human psychology is a subject of intense research in DARPA's XAI program.

\begin{figure*}[t]
     \centering
     \includegraphics[width=0.9\linewidth]{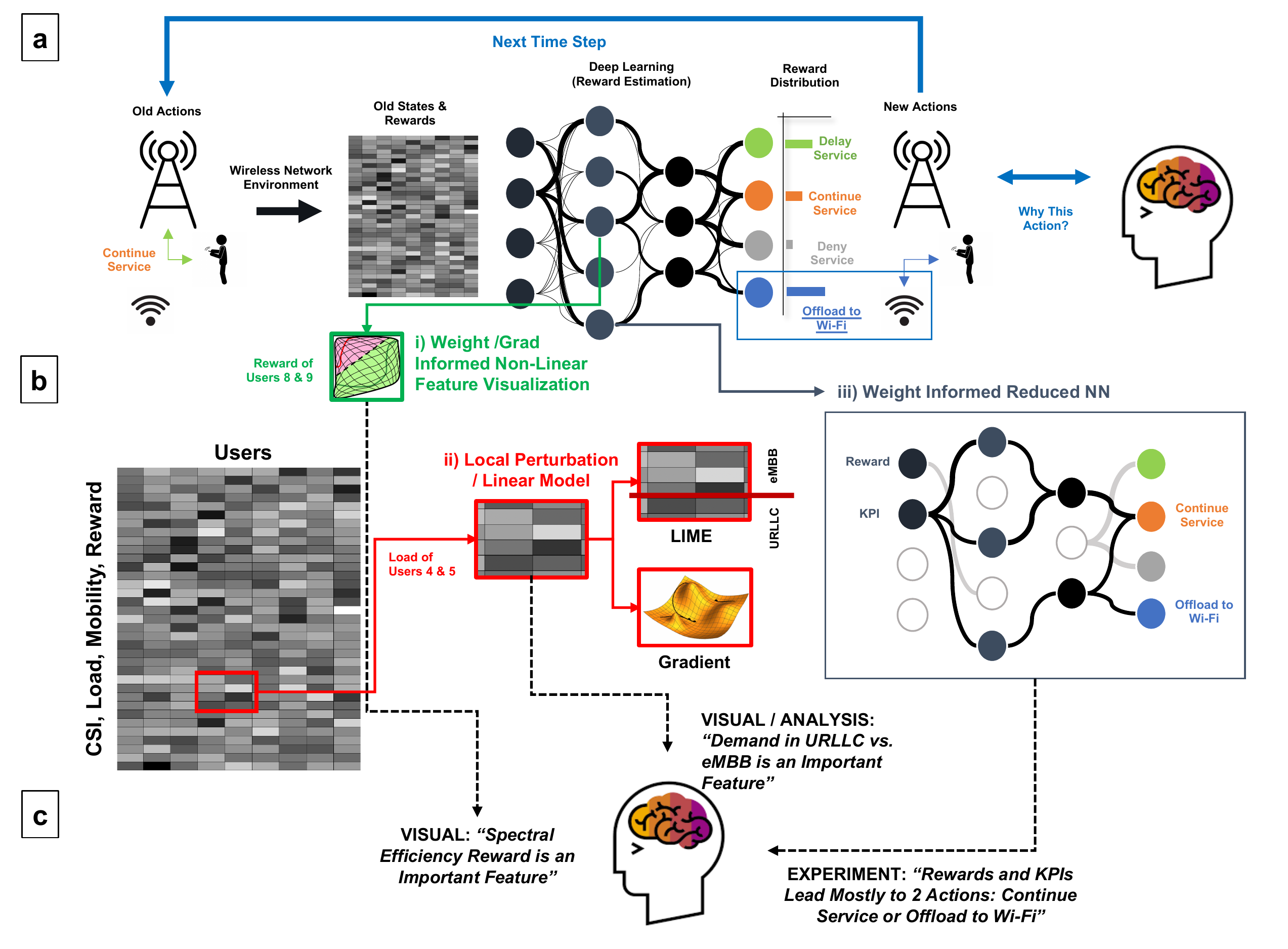}
     \caption{Explainability examples in DRL: a) DRL without explainability, b) a range of explainability options using data features and compressed neural network (NN), and c) human reasoning based on explainability.}
     \label{fig3}
\end{figure*}

\section{Methods to Improve Explainability}
\label{section4}

Here, we give a review of recent attempts to improve explainability, especially in deep learning (including DRL). We map specific methods to the main explainability capabilities we wish for in Section\ref{section2}, which were: 1) Visualisation with Case Studies, 2) Hypothesis Testing, and 3) Didactic Statements. 

\subsection{Physics Informed Design}

Designing DL algorithms that are physics based can negate many of the concerns, as they have direct explainability. For example, equalising the nonlinear channel loss (e.g. a multitude of dispersion and phase noise in NLSE channels) is traditionally achieved via digital back propagation methods such as Split-Step Fourier Method (SSFM). Designing DNN that approximates this process in the form of a Learned Digital Back Propagation (LDBP) is achieved by unrolling the SSFM iterations and approximating each span inversion with 2 layers \cite{8386120}. However, in many cases, this is not possible because we lack a workable traditional model or that it has unsatisfactory performance.

\subsection{Visualisation Techniques}

At the perhaps most intuitive level of explainability, one can visualise the features that are important based on their weights or gradients of local nodes in the NN. In a gradient based approach, we calculate the gradient of each input feature with respect to an output:
\begin{equation}
f'(x) = \lim_{\Delta x \rightarrow 0} \frac{(x+\Delta x)-x}{\Delta x},
\end{equation} where a small change in the input data feature leads to the level of outcome change can be visualised. An example of visual outputs in Fig.~\ref{fig3}b-i include the spectral efficiency (SE) reward of users 8 \& 9 and its high impact on the output actions. Local features in hidden layers are non-linear and therefore the interpretation maybe not trivial. This explainability process can be further enhanced by yielding didactic statement explanations by layer-wise relevance propagation (reversing the NN by weight importance).  

\subsection{Local Data and Local Model Reduction}
Instead of reducing the global DL model, we can also create simpler surrogate models of selected partial data. For example, we can select only the load demand data (see states in Fig.~\ref{fig3}b-ii) to see how this input feature affects the output. In general, let the model being explained be $f$, then one attempts to identify one or a set of interpretable model $g \in G$ (such as the interpretable linear models, decision trees, rule tables discussed previously) that is locally faithful to the classifier in question: $\hat{y} = g(x^{*})$, where $x^{*}$ is a subset of $x$ \cite{XAI2}. We can also create local explainable models (e.g. local linear model $\hat{y} = \theta x^{*} + n$) to understand better what DL is doing. In Fig.~\ref{fig3}b-ii, we can see that the load of users 4 \& 5 influence action choice and can be local linearly divided between the URLLC and eMBB load demand - and this output can be either visual or quantitative analysis. One popular approach based on the above logic is called Local interpretable model-agnostic explanations (LIME) \cite{XAI-review}. LIME introduces a measure of complexity for $g$: $\Omega(g)$; such that one solves the following to obtain the minimum explanation $\xi(x)$:
\begin{equation}
    \label{LIME}
    \xi(x) = \underset{g\in G}{\arg\max} \quad \mathcal{L}(f,g,\Sigma_x) + \Omega(g),
\end{equation} where $\mathcal{L}(f,g,\Sigma_x)$ is a measure of how unfaithful reduced model $g$ is in approximating $g$ in the locality of $\Sigma_x$. As such, LIME quantifies the simplest explanation by minimizing the error of local model reduction and its complexity.

\subsection{Global Model Reduction Techniques}
Since we know that simpler models are more likely to be explainable, e.g. fewer parts to link mathematically, more likely to be in a form we recognise, ...etc., and as such model reduction makes sense. There are a multitude of ways in which this can be achieved with varying results and we detail some, but not all approaches below.

\subsubsection{Problem Reduction}
In reinforcement learning, the framework is often formulated from a Markov Decision Process (MDP). The size of MDP is directly determined by the state and action spaces, which grow super-polynomially with the number of variables that characterise the domain. To support fine-grained RRM, we have to adopt high-resolution communication context to accommodate context-aware optimization, which often results in a large-scale Partially Observable MDP (POMDP). The worst-case complexity is determined by the model, ranging from POMDP with PSPACE-complete (polynomial to input) to PO Stochastic Games with NEXP-complete (non-deterministic Turing machine using time $2^{n^{\mathcal{O}(1)}}$) complexity. In general, one can compress MDP model in two stages:
\begin{itemize}
	\item MDP model construction: one can appropriately choose the definitions of state and/or action to adjust their resolution. For example, when the transmit power constitutes the action space, we could use a limited number of discretised levels to approximate their dynamic range with controlled performance loss. Example: hierarchical action space methods can be used to approximate the POMDP problem, achieving a scalable compression. 
	\item During the learning process: the size of MDP model can be further reduced by aggregating identical or similar states, allowing us to reduce learning complexity with a bounded loss of optimality \cite{state_abstraction}. The similarity of states can be measured in terms of optimal Q function, reward and state transitions, Boltzmann distributions on Q values, \etc. 
\end{itemize}

\subsubsection{Neural Network Reduction}
Previous studies have revealed that NNs are typically over parameterised \cite{sparsity}, and one can achieve similar function approximation by removing components (e.g. pruning the network as shown in Fig.~\ref{fig3}b-iii) and only retaining useful parts with greatly reduced model size. There are several typical ways on compressing DNN by exploiting sparsity in NN:
\begin{itemize}
    \item Reducing the number of parameters: removing the number of connections/weights, or pruning filters. 
    \item Architectural reform: replacing fully-connected layers with more compact convolutional layers. 
    \item Quantization: reduce the bit width integer to store weights. 
\end{itemize}

In general, selecting appropriate local data or reducing the global model also gives extra explainability power by developing experiential and example based explanations, including testing hypothesis (e.g. does a selected feature set cause the outcome we expect from traditional wisdom?).

\begin{figure}[t]
     \centering
     \includegraphics[width=1.0\linewidth]{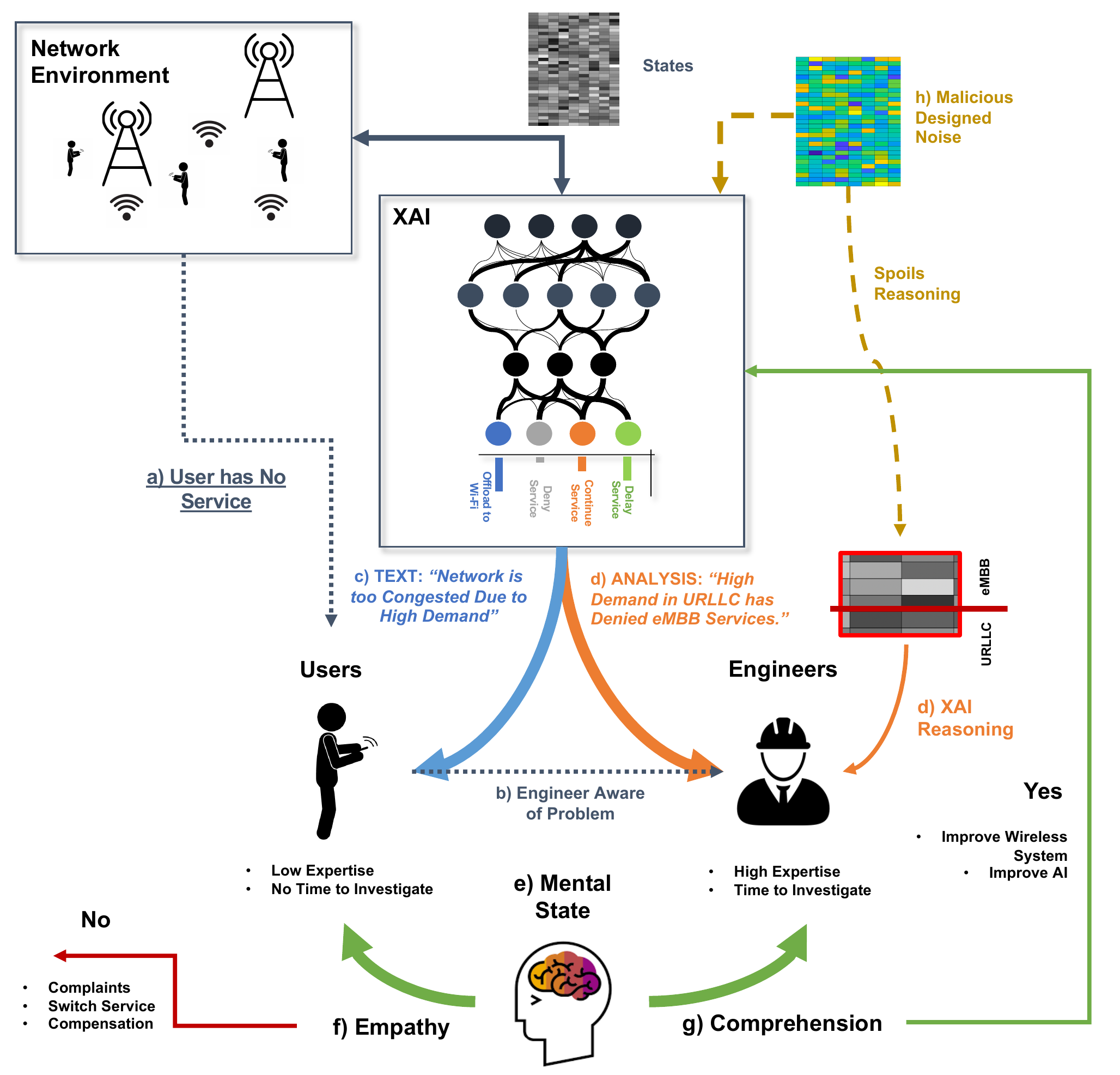}
     \caption{XAI Integration Framework in Future Wireless Networks: XAI provides diverse explanations to consumer users and radio engineers with varying levels of expertise and demand, their mental state affects their interpretation and this feeds back to the ecosystem. Challenges include malicious attacks using designer noise.}
     \label{fig5}
\end{figure}

\section{XAI Integration into 6G: Framework and Future Challenges}
\label{section5}

\subsection{Framework and Open Challenges}
In the context of Beyond 5G and 6G, the main areas that require improved trust are mainly in automation: 1) transport, 2) precision manufacturing, 3) healthcare, and 4) human machine brain interface. The framework we propose in Fig.~\ref{fig5} is a general one which explains to users and radio engineers the current actions of the network. In our example scenario, a) service is denied to users that b) the engineer is aware of through classical mechanisms of network monitoring and diagnostics. XAI provides differing levels of explanations to them in order to achieve different goals: c) user is given a simple didactic statements via APP/Social Media/brain interface, and d) engineer is given quantitative analysis via local linear model classification of a key feature set. Their e) mental states affect their understanding level and cognitive bias. For f) users - empathy is more important to maintain high customer satisfaction, and failure to do so can lead to complaints, compensation, and churn for the network. For g) engineers - comprehension is critical to affect change in the network/AI if needed.

The open challenges are numerous and we list the following two multi-disciplinary areas:
\begin{enumerate}
    \item \textbf{Human Machine (Brain) Interface:} developing rational and intuitive interfaces (proprietary or existing) that communicate (e.g. didactic statements, interactive visual, brain wave) to users and engineers - without disrupting their lives and fit into existing workflows and processes. In particular, it needs to tackle the cognitive biases in human minds, their mental states, which all impact on their degree of comprehension and empathy. The recent advances in human-brain interfacing \cite{6G} for tactile control and shared intelligence presents a futuristic framework for XAI. 
    \item \textbf{XAI Twin:} develop an explainable twin AI system to work in parallel to the deep learning systems that are designed for optimisation performance. The XAI twin enables us not to sacrifice performance trade-off (as seen in Fig.~\ref{fig2}, whilst offering intuitive explanations. Recent work to develop a Neuro-Symbolic Concept Learner (NS-CL) agent that mimics human concept learning, able to translate back to the language description of the features \cite{ICLR19}.
    \item \textbf{Defence against Attacks:} We saw in the example given in Fig~\ref{fig5}h \cite{ICLR14} that a targeted noise input (with no clear features) can lead to catastrophic errors to a well trained DL engine. How can we develop defence mechanisms that can recognise targeted attacks against DL and XAI engines? 
\end{enumerate}

\subsection{Conclusions}

As 6G will need to enable greater levels of autonomy across a wide range of industries, building trust between human end users and the enabling AI algorithms is critical. At the moment, we simply don't understand a wide range of deep learning modules that contribute to PHY and MAC layer roles, ranging from channel estimation to cross-RAT access optimisation. The need for increased explainability to enable trust is critical for 6G as it manages a wide range of mission/safety critical services as well as interfacing human brain and machines directly. In this review, we outlined the core concepts of Explainable Artificial Intelligence (XAI) for 6G, including: public and legal motivations, definitions of explainability, performance vs. explainability trade-offs, methods to improve explainability, and proposed a framework to incorporate XAI into future wireless systems. Our review has been grounded in cases studies for both PHY and MAC layer optimisation, and provide the community with an important research area to embark upon. \\

\textbf{Acknowledgements:} The author wishes to acknowledge EC H2020 grant 778305: DAWN4IoE - Data Aware Wireless Network for Internet-of-Everything, and The Alan Turing Institute under the EPSRC grant EP/N510129/1.

\bibliographystyle{IEEEtran}
\bibliography{Ref}

\begin{thebibliography}{10}
\providecommand{\url}[1]{#1}
\csname url@samestyle\endcsname
\providecommand{\newblock}{\relax}
\providecommand{\bibinfo}[2]{#2}
\providecommand{\BIBentrySTDinterwordspacing}{\spaceskip=0pt\relax}
\providecommand{\BIBentryALTinterwordstretchfactor}{4}
\providecommand{\BIBentryALTinterwordspacing}{\spaceskip=\fontdimen2\font plus
\BIBentryALTinterwordstretchfactor\fontdimen3\font minus
  \fontdimen4\font\relax}
\providecommand{\BIBforeignlanguage}[2]{{%
\expandafter\ifx\csname l@#1\endcsname\relax
\typeout{** WARNING: IEEEtran.bst: No hyphenation pattern has been}%
\typeout{** loaded for the language `#1'. Using the pattern for}%
\typeout{** the default language instead.}%
\else
\language=\csname l@#1\endcsname
\fi
#2}}
\providecommand{\BIBdecl}{\relax}
\BIBdecl

\bibitem{6G}
W.~Saad, M.~Bennis, and M.~Chen, ``{A Vision of 6G Wireless Systems:
  Applications, Trends, Technologies, and Open Research Problems},'' \emph{IEEE
  Network}, 2020.

\bibitem{AI5G}
R.~{Li}, Z.~{Zhao}, X.~{Zhou}, G.~{Ding}, Y.~{Chen}, Z.~{Wang}, and H.~{Zhang},
  ``{Intelligent 5G: When Cellular Networks Meet Artificial Intelligence},''
  \emph{IEEE Wireless Communications}, vol.~24, no.~5, pp. 175--183, October
  2017.

\bibitem{DRL_application_survey}
N.~C. {Luong}, D.~T. {Hoang}, S.~{Gong}, D.~{Niyato}, P.~{Wang}, Y.~{Liang},
  and D.~I. {Kim}, ``{Applications of Deep Reinforcement Learning in
  Communications and Networking: A Survey},'' \emph{IEEE Communications Surveys
  Tutorials}, pp. 1--1, 2019.

\bibitem{UAV-AI3}
U.~{Challita}, W.~{Saad}, and C.~{Bettstetter}, ``{Interference Management for
  Cellular-Connected UAVs: A Deep Reinforcement Learning Approach},''
  \emph{IEEE Transactions on Wireless Communications}, vol.~18, no.~4, pp.
  2125--2140, April 2019.

\bibitem{Li18}
H.~{Ye}, G.~Y. {Li}, and B.~{Juang}, ``{Power of Deep Learning for Channel
  Estimation and Signal Detection in OFDM Systems},'' \emph{IEEE Wireless
  Communications Letters}, vol.~7, no.~1, pp. 114--117, Feb 2018.

\bibitem{Farsad18}
N.~{Farsad} and A.~{Goldsmith}, ``Neural network detection of data sequences in
  communication systems,'' \emph{IEEE Transactions on Signal Processing},
  vol.~66, no.~21, pp. 5663--5678, Nov 2018.

\bibitem{DRL_survey}
K.~Arulkumaran, M.~P. Deisenroth, M.~Brundage, and A.~A. Bharath, ``Deep
  reinforcement learning: A brief survey,'' \emph{IEEE Signal Processing
  Magazine}, vol.~34, no.~6, pp. 26--38, Nov 2017.

\bibitem{Brittle}
H.~Owhadi, C.~Scovel, and T.~Sullivan, ``On the brittleness of bayesian
  inference,'' \emph{SIAM Review}, vol.~57, no.~4, p. 566–582, April 2015.

\bibitem{ICLR14}
C.~Szegedy, W.~Zaremba, I.~Sutskever, J.~Bruna, D.~Erhan, I.~Goodfellow, and
  R.~Fergus, ``Intriguing properties of neural networks,'' in
  \emph{International Conference on Learning Representations (ICLR)}, 2014, pp.
  1--10.

\bibitem{8386120}
C.~{Häger} and H.~D. {Pfister}, ``Nonlinear interference mitigation via deep
  neural networks,'' in \emph{2018 Optical Fiber Communications Conference and
  Exposition (OFC)}, March 2018, pp. 1--3.

\bibitem{XAI2}
M.~T. Ribeiro, S.~Singh, and C.~Guestrin, ``"why should i trust you?":
  Explaining the predictions of any classifier,'' in \emph{ACM SIGKDD
  International Conference on Knowledge Discovery and Data Mining}.\hskip 1em
  plus 0.5em minus 0.4em\relax New York, NY, USA: ACM, 2016, pp. 1135--1144.

\bibitem{XAI-review}
S.~{Chakraborty}, R.~{Tomsett}, R.~{Raghavendra}, D.~{Harborne}, M.~{Alzantot},
  F.~{Cerutti}, M.~{Srivastava}, A.~{Preece}, S.~{Julier}, R.~M. {Rao}, T.~D.
  {Kelley}, D.~{Braines}, M.~{Sensoy}, C.~J. {Willis}, and P.~{Gurram},
  ``Interpretability of deep learning models: A survey of results,'' in
  \emph{IEEE SmartWorld}, Aug 2017, pp. 1--6.

\bibitem{state_abstraction}
D.~Abel, D.~Hershkowitz, and M.~Littman, ``Near optimal behavior via
  approximate state abstraction,'' in \emph{ICML}, ser. Proceedings of Machine
  Learning Research, vol.~48, Jun 2016, pp. 2915--2923.

\bibitem{sparsity}
W.~Wen, C.~Wu, Y.~Wang, Y.~Chen, and H.~Li, ``Learning structured sparsity in
  deep neural networks,'' in \emph{NIPS}, Dec 2016, pp. 2082--2090.

\bibitem{ICLR19}
J.~Mao, C.~Gan, P.~Kohli, J.~Tenenbaum, and J.~Wu, ``The neuro-symbolic concept
  learner: Interpreting scenes, words, and sentences from natural
  supervision,'' in \emph{International Conference on Learning Representations
  (ICLR)}, 2019, pp. 1--10.

\end{thebibliography}

\end{document}